\documentstyle[aps,preprint,prb,epsf]{revtex}
\tightenlines

\begin{document}

\title{Power law correlations in the Southern Oscillation Index
fluctuations characterizing El Ni$\tilde{n}$o}

\author{ M. Ausloos$^{1}$ and K. Ivanova$^{2}$}
\address{$^{1}$ SUPRAS and GRASP, B5,
Sart Tilman Campus, B-4000 Li$\grave e$ge, Belgium,\\
$^{2}$ Department of
Meteorology, Pennsylvania State University, University Park, PA 16802, USA}

\date{\today} 

\draft

\maketitle

\begin{abstract} {The southern oscillation index ($SOI$) is a
characteristic of
the El Ni$\tilde{n}$o phenomenon. $SOI$ monthly averaged data is analyzed
for the
time interval 1866-2000. The tail of the cumulative distribution of the
fluctuations of $SOI$ signal is studied in order to characterize the amplitude
scaling of the fluctuations and the occurrence of extreme events. Large
fluctuations are more likely to occur than the Gaussian distribution would
predict. The time scaling of fluctuations is studied by applying the energy
spectrum and the Detrended Fluctuation Analysis ($DFA$) statistical method.
Self-affine properties are found to be pertinent to the $SOI$ signal and
therefore suggest power law correlations of fluctuations of the signal.
Antipersistent type of correlations exist for a time interval ranging
from about
4 months to about 6 years. This leads to favor specific physical models for El
Ni$\tilde{n}$o description.}

\end{abstract}

\vskip 1cm

El Ni$\tilde{n}$o is one of the most fascinating phenomena in meteorology. Its
consequences are thought to be wide ranged and sometimes catastrophic.
There is
no fundamental explanation, even though much work has been performed on the
subject.\cite{websites} One basic function characterizing el Ni$\tilde{n}$o is
the so-called {\it southern oscillation index} ($SOI$). The cyclic warming and
cooling of the eastern and central regions in the Pacific Ocean are coupled to
distinctive sea level pressures. The normalized difference between the
pressure
measured at Darwin and the pressure measured at Tahiti are used to generate
the
so-called $SOI$ number. When the number is negative there is an
El-Ni$\tilde{n}$o
(or ocean warming), but when the number is positive, there is a
La-Ni$\tilde{n}$a
effect (or ocean cooling). Indeed the central and eastern Pacific
regions of the
ocean are normally colder than its equatorial location would suggest,
mainly due
to the influence of northeasternly trade winds, a cold ocean current
flowing up
the coast of Chile, and to the up-welling of cold deep water off the coast of
Peru. At times, the influence of these cold water sources wane, causing the
surface of the eastern and central Pacific to warm up under the tropical sun.

There is already a large number of $SOI$
analyses.\cite{nicolis2,hasselmann1,fraedrich1,palmer} Here we analyze the
southern oscillation index monthly averaged data for the time
interval 1866-2000,
i.e. 1612 data points.\cite{data1,data2} Even though this is a
limited number of
data points, it is possible to study the amplitude and time scaling of the
fluctuations of the $SOI$ data by calculating the power spectrum and
applying the
Detrended Fluctuation Analysis statistical method ($DFA$)\cite{DFA}
with acceptable error bars. The latter
technique has already demonstrated its usefulness in
turbulence,\cite{Ghash} in biology for
sorting out coding and non-coding sequence behavior in DNA,\cite{DNA} in
financial analysis for characterizing e.g. currency exchange
rates,\cite{nvma,kiandma1} and in other self-organizing critical
systems\cite{Bak} for time series with similar lengths.

We do not aim at forecasting\cite{forecasts} here. Forecasting models already
exist for El-Ni$\tilde{n}$o events, through the sea-surface
temperature canonical
correlation analysis ($CCA$) model,\cite{barnett,cca} the coupled
ocean/atmosphere model,\cite{coam} and the linear inverse model\cite{lim} to
mention a few of them.

Note also, an excellent, detailed intercomparison of three different
forecasting
models that have predicted moderate warming of the equatorial Pacific  sea
surface temperatures (SSTs) to begin in late summer to early fall of
1986.\cite{barnett} One of the models includes sea level pressure data, a
quantity that is subject to our study. This is a statistical model, that is
designed to predict sea surface temperature from {\it prior} variations of the
sea level pressure. Authors draw a conclusion that the key to the
success of the
model is recognizing the large-scale, low-frequency changes in the tropical
ocean-atmospheric system. They based their conclusion on the estimate of the
relative importance of the {\it prior data}, included in the model, for the
subsequent predictive skill of the model.

The present work stresses the implications of correlations in the amplitude
and
time scaling of the fluctuations of $SOI$. In so doing, a selected
choice between
proposed models can be envisaged.

\section{Data}

Data for years 1866-2000 were obtained from Ref.\onlinecite{data1} and
Ref.\onlinecite{data2}. They are plotted as function of time (in
month units) in
Fig. 1. The data consists of $1612$ data points and represent monthly averaged
normalized difference between the pressure measured at Darwin and the sea
level
pressure measured at Tahiti. For the years before 1866 daily
measurements of the
sea level pressure at both stations have been reported to exist and monthly
values of the southern oscillation index have been calculated \cite{jones}
back
to 1841. However, there are gaps of a couple of years in the record whence not
suitable for our analysis.

\section{Distribution of fluctuations}

First, we calculate the fluctuations $y_{i+1}-y_i$ of the $SOI$ signal $y(t)$,
group them into bins with size equal to the difference between the maximum and the minimum
of the fluctuations and
divide by one hundred, i.e. $~0.082$. Then we count the number of
entries inside
each bin. The result is the histogram shown in the inset of Fig. 2. The
distribution of fluctuations of the $SOI$ signal is symmetrical and
not Gaussian.

Next we focus on the tails as they characterize how likely the
extreme events are
to occur.  We calculate the empirical probability to observe
fluctuations with an
amplitude larger than some value, $Prob(X>x)$, where $x=|y_{i+1}-y_i|$. The
result is plotted in Fig. 2 (dots). The tail of this distribution is
consistent
with a power law $Prob(X>x) \sim x^{-\mu}$, showing the reduction of the
probability for increasing intensity of the fluctuations. A linear
least-squares
fit yields an estimate $\mu=3.30 \pm 0.06$ when the amplitudes of fluctuations
are between 1.5 and 2.8. It is difficult to have good estimate for larger
amplitudes of the fluctuations (between 3 and 4) due to insufficient
statistics.
Note, that the $\mu$-value is well outside the range for stable L\'evy
distributions, $0<\mu<2$, and in particular outside the Gaussian distribution
($\mu=2$). Therefore, large fluctuations are more likely to occur than the
Gaussian distribution would predict.

The cumulative probability distribution provides an estimate for the intensity
structure, e.g. the amplitude scaling of the fluctuations of the $SOI$ signal.
The asymptotic power law behavior of the intensity distribution of the
fluctuations appears to be a particularly suitable description of the
occurrence
of extreme events. In the next section we will focus on the time scaling and
correlations of the fluctuations of the $SOI$ signal.

\section{$SOI$ temporal correlations}

We apply two methods to study the temporal correlations of fluctuations in the
$SOI$ signal, spectral and the detrended fluctuation analysis methods.

The spectral analysis involves calculating of the power spectrum as a Fourier
transform of the data. To quantify the correlations of the fluctuations in a
signal the scaling properties of its power spectrum are tested. Assuming a
power-law behavior of the energy spectrum $S(f) \sim f ^{-\beta}$ the
self-affine
properties of the signal are characterized by the $\beta$-value.

In Fig. 3 results from the spectral analysis of the $SOI$ data are plotted.
For
the frequency range from 1/5 to 1/64 months$^{-1}$ the spectrum is consistent
with a power-law behavior with a spectral exponent $\beta=1.32 \pm
0.14$. However
such Fourier transform analyses fall short of precisely showing crossover
regimes. To better estimate the crossover and to test the correlations using a
different approach, we analyze the $SOI$ signal applying the $DFA$
technique.\cite{DFA}

The detrended fluctuation function $F(\tau)$ is calculated following

\begin{equation} {{F^2(\tau)}} = {1 \over \tau} {\sum_{t=k\tau+1}^{(k+1)\tau}
{\left[y(t)-z(t)\right]}^2} , {\hskip 1cm} k=0,1,2, \cdots, ({N \over
\tau} -1),
\end{equation}

\noindent where $z(t)=at+b$ is a linear least-square fit to the data
points in a
box containing $\tau$ points.

The behavior of the averaged though other forms for $z(t)$ can be used \cite{madub}
$F^2(\tau)$ over the $N/\tau$ intervals
with length
$\tau$ is expected to be a power law

\begin{equation} \langle F^2(\tau) \rangle \sim \tau^{2\alpha}. \end{equation}

{\noindent  An exponent $\alpha \not= 1/2$ in a certain range of $\tau$ values
implies the existence of long-range correlations in that time interval as, for
example, in fractional Brownian motion \cite{west,brown,Addison}.  A value
of $\alpha <
0.5$ indicates antipersistence and $\alpha > 0.5$ implies persistence of
correlations. The classical random walk (Brownian motion) is such that
$\alpha=1/2$.

In Fig. 4, a log-log plot of the function $\sqrt{\langle F^2(\tau)
\rangle}$ is
shown for the $SOI$ data in Fig. 1. This function is close to a power
law with an
exponent $\alpha_1 = 0.25 \pm 0.01$ holding for the interval time ranging from
about 4 to 70 months. In contrast to the Fourier spectrum analysis
the crossover
of the fluctuations function $<F^2(\tau)>$ is well defined at 70
months. For time
scales longer than 70 months, i.e. about 6 years, a crossover to noise-like
$\alpha_2 = 0.05\pm 0.009$ is observed. This suggests antipersistence of the
correlations in the fluctuations of the sea level pressure for time lags less
than 6 years. Antipersistence of the fluctuations implies that a positive
fluctuation in the past is more likely to be followed by a negative
fluctuation
in the future.

Note that $\alpha=Hu$, $Hu$ being the so called Hurst exponent. The Hurst
exponent of a signal was first defined in the ``rescaled range (R/S)
analysis''
(of Hurst~\cite{Hu4,feder}) to estimate the correlations in the Nile floodings
and droughts. The relationship

\begin{equation} \beta = 2 Hu + 1 \end{equation}

\noindent has been theoretically proven by Flandrin for fractional Brownian
walks.\cite{flandrin} The $\alpha_1$ value for the $SOI$ signal is consistent
within the error bars with the spectral exponent satisfying the above
equation.

\section{Conclusions}

We have studied the amplitude, time scaling and correlations of the
fluctuations
of the Southern Oscillation Index ($SOI$).  The tail of the cumulative
distribution of the $SOI$ fluctuations is found to scale with an exponent
$\mu=3.3$ for amplitudes of the fluctuations between 1.5 and 2.8,
describing the
occurrence of extreme events. The energy spectrum of the $SOI$  is consistent
with power law with exponent $\beta=1.32$ for the frequency range
between 1/5 and
about 1/64 months. Since $1<\beta<3$ the $SOI$ signal is a self-affine
fractal.\cite{brown}
To estimate more precisely the crossover regime and the type of
correlations we
have applied the $DFA$ method. Using the $DFA$ method we find an
antipersistent
type of correlations for time lags less than 70 months. The $\alpha$-exponent
that characterizes the scaling is consistent within the error bars with the
spectrum scaling. This leads to favor specific physical models for El
Ni$\tilde{n}$o description, as that in Ref.\onlinecite{barnett}.

Our analysis shows that long-range correlations exist between the
fluctuations of
the $SOI$, e.g. sea level pressure. This supports one conclusion in
Ref.[\onlinecite{barnett}] pertinent to our analysis, i.e. the large-scale
low-frequency variations of the global SLP field are responsible for the
predictive skill of the sea level pressure forced statistical model.
The present work indicates
a hint why it might be so: because of the correlations between the
fluctuations.

\vskip 0.6cm {\bf Acknowledgments} \vskip 0.6cm

We thank T.P. Ackerman, H.N. Shirer and E.E. Clothiaux for stimulating
discussions and comments. This investigation was partially supported by grant
number Battelle 327421-A-N4. The comments of C. Nicolis are greatly
appreciated.

\newpage 
\begin{figure}[ht] 
\begin{center} 
\leavevmode 
\epsfysize=8cm
\epsffile{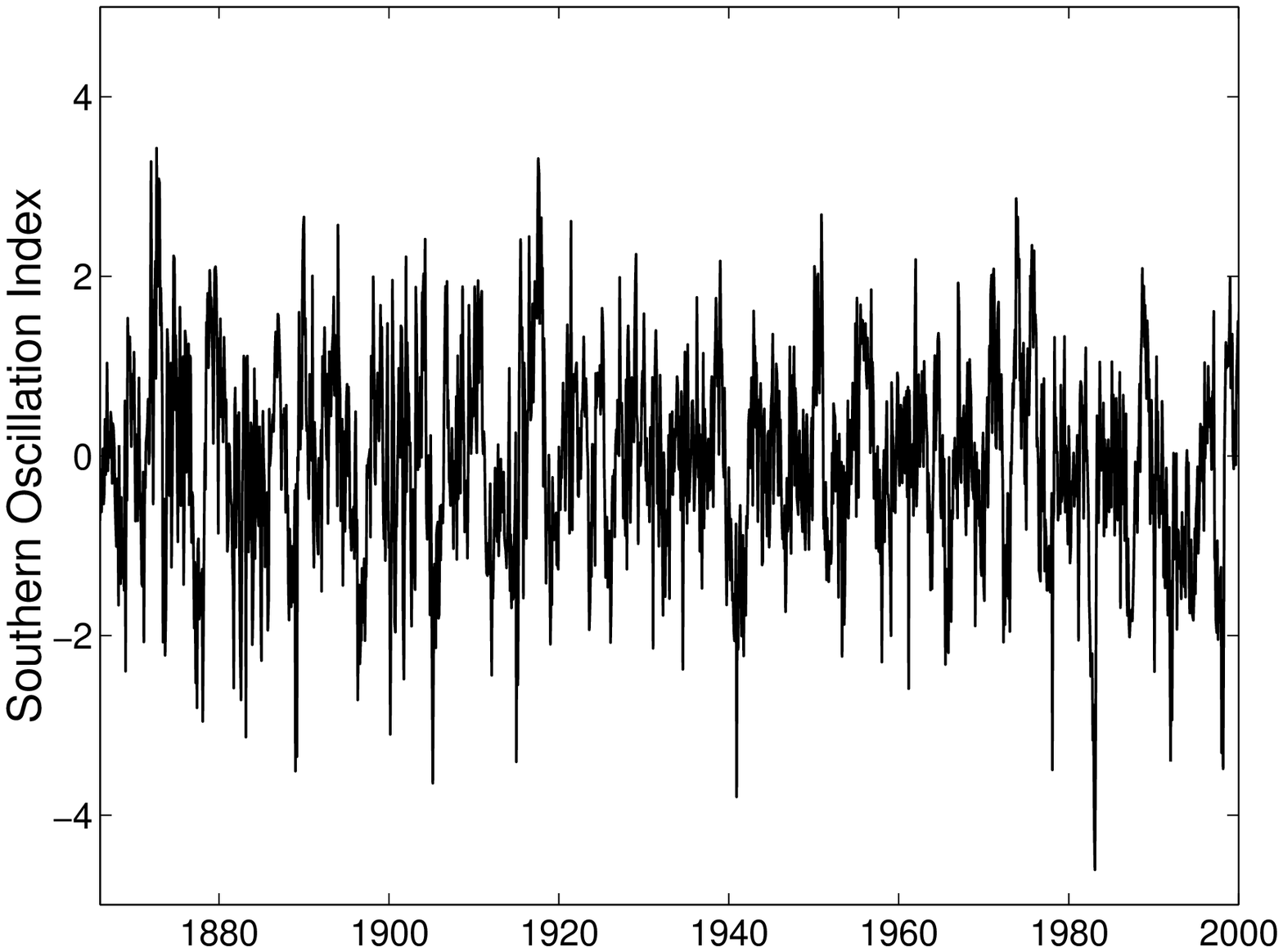} 
\end{center}
\caption{The monthly values of the Southern Oscillation Index for the
period 1866 to April 2000 representing the standardized pressure difference
between Tahiti and Darwin stations ($1612$ data points).} \label{fig1} \end{figure}

\begin{figure}[ht] 
\begin{center} 
\leavevmode 
\epsfysize=8cm
\epsffile{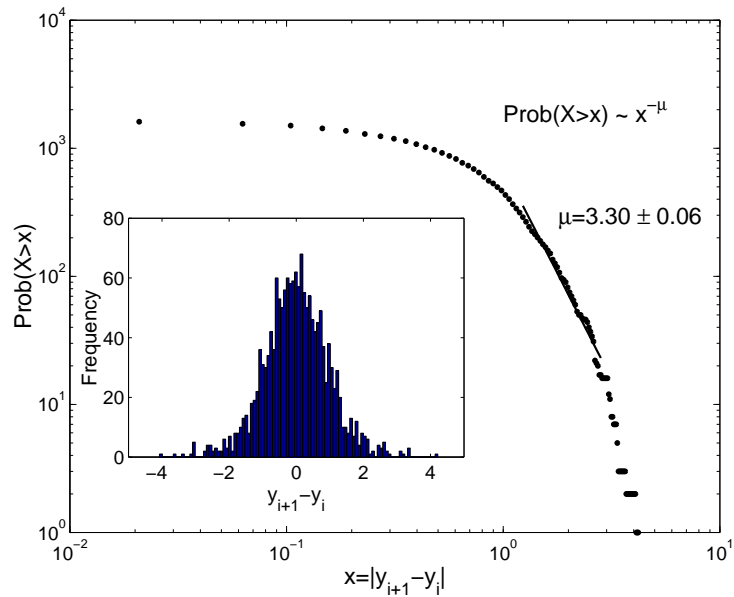} 
\end{center}
\caption{The empirical probability to observe
fluctuations
with amplitude larger than some value $x=|y_{i+1}-y_i|$. An
asymptotic power law
scaling is found for fluctuation amplitudes between 1.5 and 2.8.
Histogram of the
fluctuations of the $SOI$ signal is shown in the inset.} \label{fig2} \end{figure}

\begin{figure}[ht] 
\begin{center} 
\leavevmode 
\epsfysize=8cm
\epsffile{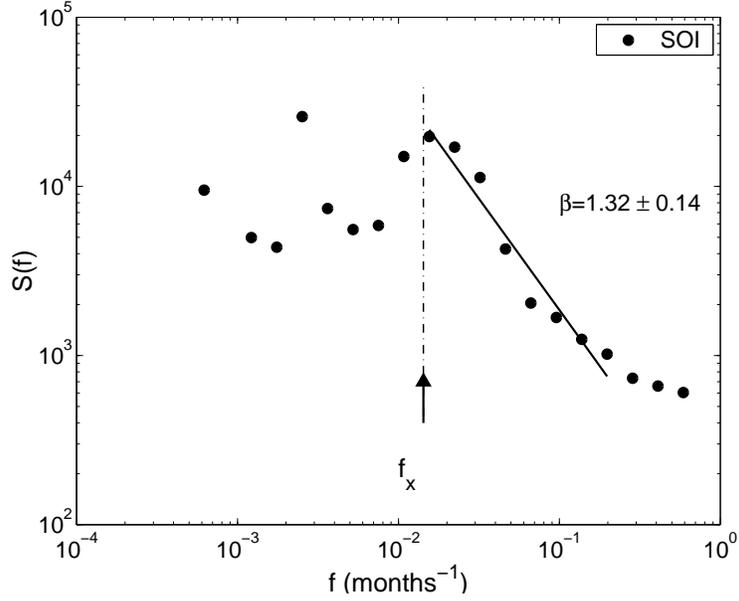} 
\end{center}
\caption{The energy spectrum of the $SOI$ data (from
Fig.
1). A spectral exponent $\beta=1.32 \pm 0.14$ characterizes the
correlations of
fluctuations in the frequency range from about 1/5 to about 1/64
months$^{-1}$.
} \label{fig3} \end{figure}

\begin{figure}[ht] 
\begin{center} 
\leavevmode 
\epsfysize=8cm
\epsffile{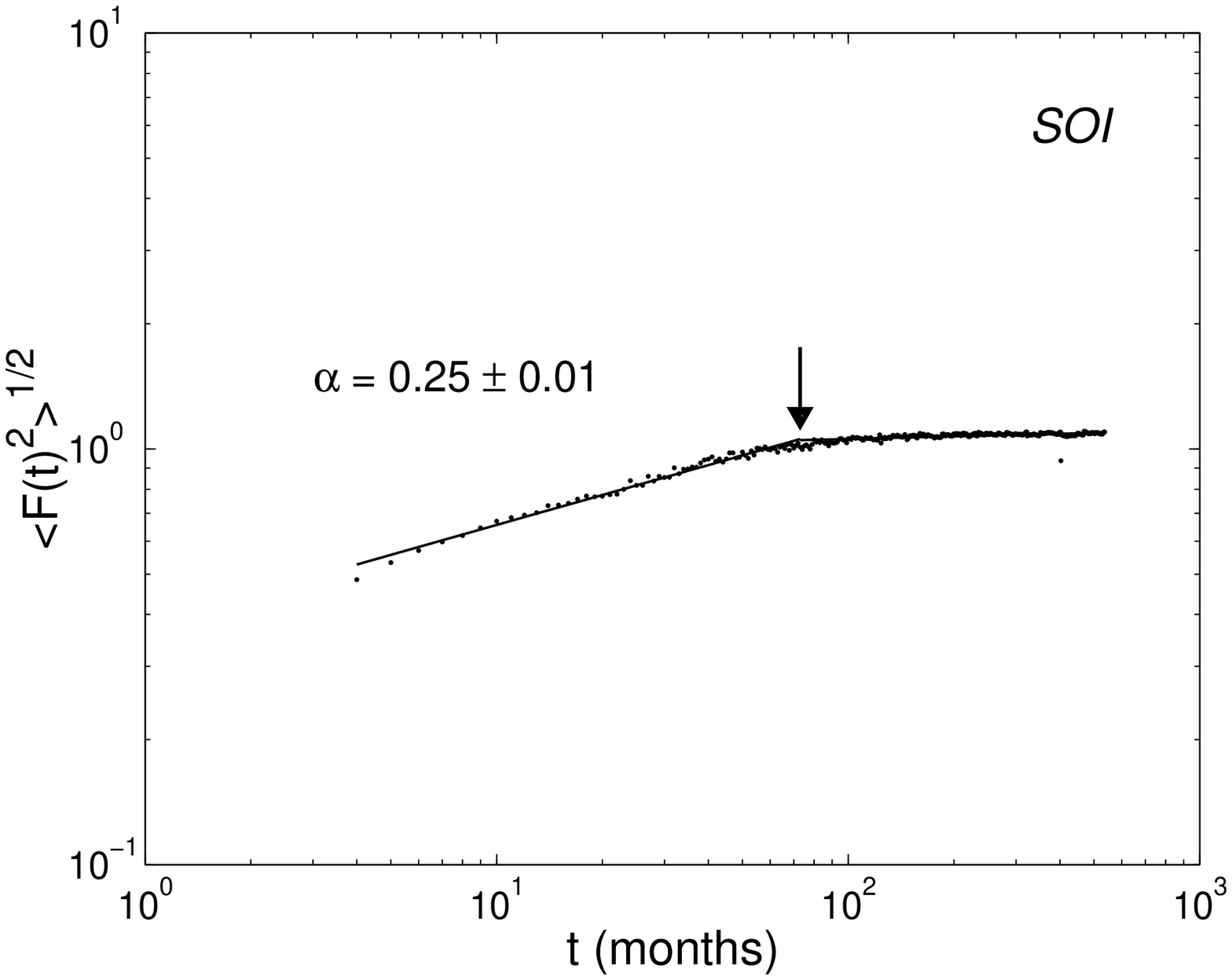} 
\end{center}
\caption{The DFA-function $<F^2(\tau)>^{1/2}$ in log-log
plot for the $SOI$ data  from Fig. 1. Two scaling regimes are
observed, $\alpha_1
= 0.25 \pm 0.01$ and noise-like $\alpha_2 = 0.05\pm 0.009$ with a
crossover at 70
months.} \label{fig4} \end{figure}


\vskip 1cm

\begin{thebibliography}{99}

\bibitem{websites} There are several websites where information and data are
available; see e.g.

http://www.pmel.noaa.gov/toga-tao/el-nino/forecasts.html

http://www.cpc.ncep.noaa.gov/

http://ingrid.ldgo.columbia.edu/SOURCES/.Indices/ensomonitor.html

$http://www-jisao.atmos.washington.edu/data_sets/$

\bibitem{nicolis2} C. Nicolis, {\it Tellus}, {\bf 42A}, 401-412 (1990).

\bibitem{hasselmann1} C. Frankignoul and K. Hasselmann, {\it Tellus}, {\bf 29}
289-305 (1977).

\bibitem{fraedrich1} K. Fraedrich, {\it Monthly  Weather Review},
{\bf 116} 1001
(1988).

\bibitem{palmer} T.N. Palmer {\it Rep. Prog. Phys.}, {\bf 63} 71 (2000).

\bibitem{data1} Thanks to James Mather from University of Wisconsin
who referred
us to:

$http://www-jisao.atmos.washington.edu/data_sets/$

\bibitem{data2} http://www.cpc.ncep.noaa.gov/data/indices/

\bibitem{DFA} C.-K. Peng, S.V. Buldyrev, S. Havlin, M. Simons, H.E.
Stanley, and
A.L. Goldberger, {\it Phys. Rev. E} {\bf 49}, 1685 (1994).

\bibitem{Ghash} S. Ghashghaie, W. Breymann, J. Peinke, P. Talkner and Y.
Dodge,
{\it Nature} {\bf 381}, 767 (1996).

\bibitem{DNA}  H.E. Stanley, S.V. Buldyrev, A.L. Goldberger, S. Havlin, C.-K.
Peng, and M. Simons, {\it Physica A} {\bf 200}, 4 (1996).

\bibitem{nvma} N. Vandewalle and M. Ausloos, {\it Physica A} {\bf 246}, 454
(1997).

\bibitem{kiandma1} K. Ivanova and M. Ausloos, {\it Physica A} {\bf 270}, 526
(1999).

\bibitem{Bak} P. Bak, K. Chen and M. Creutz, {\it Nature} {\bf 342},
780 (1989).

\bibitem{forecasts}
$http://www.cpc.ncep.noaa.gov/products/analysis_monitoring/bulletin/
forecast.html$


\bibitem{barnett} T. Barnett, N. Graham, M. Cane, S. Zebiak, S. Dolan, J.
O'Brien, and D. Legler, {\it Science}, {\bf 241} 192 (1988)

\bibitem{cca} A.G. Barnston and C.F. Ropelewski, {\it Journal of Climate},
{\bf
5}, 1316 (1992).

\bibitem{coam} M. Ji, D.W. Behringer, A. Leetmaa, {\it Monthly Weather Review}
{\bf 126}, 1022 (1998).

\bibitem{lim} C. Penland and T. Magorian, {\it Journal of Climate},
{\bf 6}, 1067
(1993).

\bibitem{jones} G.P. Konnen, P.D. Jones, M.H. Kaltofen, R.J. Allan,
{\it Journal
of Climate}, {\bf 11}, 2325 (1998).


\bibitem{madub} N. Vandewalle and M. Ausloos, {\it Int. J. Comput. Anticipat.
Syst.} {\bf 1,} 342 (1998).


\bibitem{west} B.J. West and W. Deering, {\it Phys. Rep.} {\bf 246}, 1 (1994).

\bibitem{brown} J. West, {\it The Lure of Modern Science: Fractal Thinking},
(World Scient., Singapore, 1995)

\bibitem{Addison} P.S. Addison, {\it Fractals and Chaos}, (Institute
of Physics,
Bristol, 1997)

\bibitem{Hu4} H.E. Hurst, {\it Trans. Amer. Soc. Civil Eng.} {\bf 116}, 770
(1951).

\bibitem{feder} J. Feder, {\it Fractals}, (Plenum, New York, 1988)


\bibitem{flandrin} P. Flandrin, {\it IEEE Trans. Information Theory}, {\bf
35},
197 (1989).


\end{thebibliography}
\end{document}